\documentclass[letter]{aa} 

\usepackage{graphicx}
\usepackage{txfonts}
\begin{document}

   \title{Diagnostics of plasma ionisation using torsional Alf\'en waves}


   \author{I. Ballai
          }

   \institute{Plasma Dynamics Group, School of Mathematics and Statisitics, University of Sheffield,\\
              Hounsfield Road, Hicks Building, Sheffield, S3 7RH, UK\\
              \email{i.ballai@sheffield.ac.uk.}
             }

   \date{Received X; accepted X}

 
  \abstract
   {}
 {Using the recently observed torsional Alfv\'en waves in solar prominences, we determine the ionisation state of the plasma by taking into account that Alfv\'en waves propagate in a partially ionised prominence plasma. We derive the evolutionary equation of waves and compare the analytical solutions to observations to determine the number density of neutrals.}
   {Using a single fluid plasma approximation, where the wave damping is provided by the Cowling resistivity, we study the temporal evolution of waves. By comparing the solution of equations with observational data (period, amplitude, propagation speed), we determined the value of the Cowling resistivity that led us to draw a conclusion on the amount of neutrals in the partially ionised plasma, a quantity that cannot be measured directly or indirectly.}
   {Our results show that damped torsional Alfv\'en waves are an ideal diagnostic tool for the ionisation state of the plasma. Using a simple model, we find that at the observational temperature of torsional Alfv\'en waves, the number of neutrals, is of the order of $5\times 10^{10}$ cm$^{-3}$.}
   {}


   \maketitle
%

\section{Introduction}

The problem of partial ionisation in solar prominences is
one of the most fundamental questions in the generation,
evolution, and stability of these intriguing solar features.
Solar prominences rise from chromospheric plasma and are
supported at the coronal level by magnetic fields. In addition,
despite being surrounded by the million degree coronal
plasma, the magnetic field provides suitable thermal
shielding so that the temperature in prominences does not
reach a level where all the material becomes fully ionised.

The ionisation state of the plasma (the balance between charged particles and neutral atoms) is currently impossible to determine as current observational facilities are mostly based on the emission and absorption spectroscopic lines of various ions. Nevertheless, some attempts have been made to determine the presence and properties of the neutral component of the solar atmosphere. In this context, several studies have been dedicated to the measurement of the differences in ion and neutral velocities caused by the loss of collisional coupling. By simultaneously measuring the Doppler shift in ion Fe II and neutral Fe I lines over the same
volume of plasma, differences between ion and neutral velocities (decoupling) of the Evershed flow have been
found in sunspot penumbra as deep as the photosphere by Khomenko et al. (2015). Later, Khomenko et al. (2016) found non-negligable differences in He I and Ca II velocities in solar prominences. Gilbert et al. (2007) compared He I and H$\alpha$ data in multiple solar prominences in different phases of their lifecycle and detected the drainage effect across the prominence magnetic field with different timescales for He and H atom. Subsequently, de la Cruz Rodr\'igez and Socas-Navarro (2011) have reported misalignement in the visible direction of chromospheric fibrils  that were attributed (as shown later numerically by Mart\'inez-Sykora et al. 2016) to the large ambipolar diffusion, that is, when the ion-neutral collisional frequency drops, the magnetic field can slip through the neutral population. This effect is conditional upon the appearance of strong perpendicular currents. 

The diagnostic of the solar plasma has been made possible thanks to the high resolution observation of waves and oscillations in various solar structures, from sunspots to solar wind. When the observed and measured properties of waves (period, wavelength, amplitude, frequency, damping time and length) are compared to the theoretical modelling (dispersion relation, evolutionary equation), we can determine plasma and field parameters (magnitude, orientation and structure of the magnetic field, transport coefficients, heating and cooling functions, etc.) that cannot be measured directly or indirectly. The technique known as atmospheric seismology has helped scientists to make significant advances in the area of plasma and magnetic field diagnostics.

The observation of Alfv\'en waves is a rather difficult task given that these waves do not perturb density, that is, they do not generate periodic changes in line intensity. On the other hand, the problem of torsional Alfv\'en waves in solar magnetic flux tubes is a well-studied theoretical topic. The problem of torsional Alfv\'en wave generation by footpoint motion in the presence of dissipative effects has been shown to be a very effective heating mechanism even without resonant coupling to global modes (Ruderman et al. 1997). Similar results were obtained by Antolin and Shibata (2010), Copil et al. (2010), and Murawski et al. (2015). A completely new aspect of torsional Alfv\'en waves and their applicability in magnetoseismology of stratified waveguides is explored by Verth et al. (2010) and Fedun et al. (2011), who prove that torsional Alfv\'en waves are ideal tools for diagnosing the radial inhomogeneities of flux tubes, including temperature diagnostics of the plasma inside and outside the flux tube.

The use of waves and oscillations for diagnosing the ionisation degree of the plasma is not an easy task. First of all, the current observational capabilities are not suited to carry out measurements of two-fluid plasma dynamics as these often require time resolution of the order of a second or less. Single fluid descriptions (frequencies smaller than the ion-neutral collisional frequency) are possible as these coincide with the magnetohydrodynamic (MHD) range.\ However, the ionisation degree in these plasmas only enter through non-ideal effects; therefore, damping of modes can be a suitable mechanism that can provide the necessary information. 

The present Letter is intended to exploit the propagation characteristics of recently observed torsional Alfv\'en waves in the solar prominences (Kohutova et al. 2020) to infer information about the ionisation degree of the partially ionised prominence plasma. Using a very simple mathematical model to describe waves, we derived an evolutionary equation and its characteristics are compared to observed values. The modification of waves' amplitude with time is the physical phenomenon that allows us to estimate the number density of neutrals in solar prominences. 

\section{Observational background}

In a recent study, Kohutova et al (2020) report on the first detection of large-scale torsional Alfv\'en waves that propagate in the solar chromosphere. These waves were initiated by a reconnection event, which was followed by a surge of cool plasma that was previously confined in the prominence flux rope. The multi-wavelength observation of the NOAA AR 12438 event has been carried out using the Interface Region Spectrograph (IRIS) in two passbands: far-ultraviolet (FUV) and near-ultraviolet (NUV). The FUV passband is centred on 1400 $\AA$ and is dominated by two Si IV lines formed at $\log T=4.8$ in the transition region. On the other hand, the NUV passband is centred on 2796 $\AA,$ which is dominated by the Mg II K line core that formed at $\log T=4$ in the chromosphere. The active region is connected to a prominence with a twisted flux-rope structure that can be observed at the limb in AIA 304 $\AA$ and IRIS FUV and NUV channels. 

The large-scale torsional wave observed by Kohutova et al. (2020) was detected in both the imaging and spectral data. The helical motion is clearly visible in the temporal evolution of the Doppler velocity as an anti-phase oscillation at the opposite edges of the flux tube. The plasma is observed to gradually cool down following the reconnection event. These authors find that the torsional Alfv\'en wave propagates with a period of 89 s, an amplitude of 41 km s$^{-1}$, and a damping time of 136 s. The propagation speed of waves, after the projection effects were corrected, is 140 km s$^{-1}$. In what follows, we use these observed data to determine parameters of the plasma exploiting the damping of these waves.

\section{Partial ionisation diagnostics}

As explained in the Introduction, the range of waves that are possible in partially ionised prominence plasma depends on the relative value of the wave period compared to the ion-neutral collisional frequency, $\nu_{in}$, which is defined as (Braginskii 1965)
\[
\nu_{in}=\frac{\alpha_{in}}{m_in_i+m_n n_n},
\]
where 
\[
\alpha_{in}=\frac43 n_in_nm_{in}\sigma_{in}\sqrt{\frac{8k_BT}{\pi m_{in}}}
\]
is the coefficient of friction between ions and neutrals (assuming these particles have the same temperature), $m_{in}=m_im_n/(m_i+m_n)\approx m_i/2$ is the reduced mass, $\sigma_{in}=8.4\times 10^{-15}$ cm$^2$ is the collisional cross section (Vranjes and Krstic 2013), $m_i$ and $m_n$ are the ion and neutral mass, $n_i$ and $n_n$ are the ion and neutral number densities, $k_B$ is the Boltzmann constant and $T$ is the temperature. By using a FAL-3 atmospheric model (Fontenla et al. 1990) at $T=10^4$K with $n_i=2.3\times 10^{10}$ cm$^{-3}$ and $n_n=1.2\times 10^{10}$ cm$^{-3}$, we obtain a collisional frequency of $\nu_{in}\approx 92.5$ s$^{-1}$. It is worth pointing out that the collisional frequency between these large particles at this temperature has also been previously derived by Zaqarashvili et al. (2011); however, they used a collisional cross section that is two orders of magnitude less than the value used in the present study. The value used here was obtained using a quantum mechanical approach by Vranjes and Krstic (2013); therefore, it is more accurate than the classical value used by Zaqarashvili et al. (2011). According to the standard definition, collisions between ions and neutrals (so, a two fluid plasma description) are needed for frequencies larger than $\nu_{in}$ or for periods that are smaller than $1/92.5=10$ ms. For the wave periods reported by Kohutova et al. (2020),  it is obvious that the dynamics of the observed torsional Alfv\'en waves has to be described within the framework of single-fluid MHD. We should point out that since the frequencies we are dealing with are smaller than the collisional frequency between heavy particles, its expression does not appear in our equations. In reality, the value of the collisional frequency is only needed to validate the framework in which our analysis is carried out.

Since this study aims to present a technique for determining the ionisation degree of the plasma, we use a simplistic model where effects, such as gravitational stratification, plasma and field inhomogeneity, various hydrodynamic and magnetohydrodynamic transport mechanisms, are neglected in favour of resistivity, which acts along and across the equilibrium magnetic field. Accordingly, let us suppose a vertical homogeneous magnetic field (${\bf B}_0=B_0{\bf z}$) in the presence of a plasma made up of charged particles and neutrals, whose total mass density is $\rho_0$, and initially the plasma is at rest. Given the characteristics of observations, we employ a cylindrical geometry in which the symmetry axis is parallel to the ambient magnetic field, ${\bf B}_0$. The equilibrium is perturbed and the perturbation in the velocity and magnetic field are ${\bf v}=(v_r, v_{\varphi}, v_z)$ and ${\bf b}=(b_r, b_{\varphi}, b_z)$, respectively. The dynamics of the plasma in such a configuration is given by the system of linearised MHD equations
\begin{equation}
\rho\frac{\partial {\bf v}}{\partial t}=\frac{1}{\mu_0}(\nabla\times {\bf b})\times {\bf B}_0,
\label{eq:1}
\end{equation}
\begin{equation}
\frac{\partial {\bf b}}{\partial t}=\nabla \times ({\bf v}\times {\bf B}_0)+\eta\nabla^2 {\bf b}+\frac{\eta_C-\eta}{|{\bf B}_0|^2}\nabla\times\left\{\left[\left(\nabla\times {\bf b}\right)\times {\bf B}_0\right]\times {\bf B}_0\right\},
\label{eq:2}
\end{equation}
where $\mu_0$ is the permeability of free space, $\eta$ is the classical Spitzer resistivity, and $\eta_C$ is the Cowling resistivity. The two resistivity coefficients are given by (see, e.g. Ballai et al. 2019)
\begin{equation}
\eta=\frac{m_e(\nu_{ei}+\nu_{en})}{e^2n_e\mu_0}, \quad \eta_C=\eta+\frac{\xi_n^2B_0^2}{\mu_0\alpha_{n}},
\label{eq:2.1}
\end{equation}
where $m_e$ and $e$ are the electron mass and charge, $\xi_n=\rho_n/\rho_0$ is the relative mass density of neutrals compared to the total mass density, and $\nu_{ei}$ and $\nu_{en}$ are the electron-ion and electron-neutral collisional frequencies given by
\[
\nu_{ei}=\frac{n_ee^2\Lambda}{3m_e^2\epsilon_0^2}\left(\frac{m_e}{2\pi k_B T}\right)^{3/2}, \quad \nu_{en}=n_n\sigma_{en}\left(\frac{8k_BT}{\pi m_e}\right)^{1/2}.
\]
In the above expression, $\Lambda$ is the Coulomb logarithm, $\epsilon_0$ is the permeability of free space, $\sigma_{en}\approx 10^{-15}$ cm$^2$ is the electron-neutral collisional cross section, and $\alpha_n$ is the neutral friction coefficient given by
\[
\alpha_n=2 \xi_n(1-\xi_n)\frac{\rho_0^2\sigma_{in}}{m_n}\left(\frac{k_BT}{\pi m_i}\right)^{1/2}.
\]
Since we are concentrating on Alfv\'en waves that decouple from all other possible modes, we consider that $v_r=b_r=v_z=b_z=0$ and that motion is axisymmetric, meaning that $\partial/\partial \varphi=0$. To further simplify the problem, we also assume that the dissipative processes have a much longer length scale in the radial direction (compared to longitudinal direction); this means that in dissipative terms, all radial derivatives are neglected. As a result, the azimuthal components of the momentum and induction equation become
\begin{equation}
\rho_0\frac{\partial v_{\varphi}}{\partial t}=\frac{B_0}{\mu_0}\frac{\partial b_{\varphi}}{\partial z}
\label{eq:3}
\end{equation}
and
\begin{equation}
\frac{\partial b_{\varphi}}{\partial t}=B_0\frac{\partial v_{\varphi}}{\partial z}+\eta_C\frac{\partial b_{\varphi}}{\partial z^2}.
\label{eq:4}
\end{equation}
Interestingly, for the configuration that is considered, the only dissipative process that acts upon torsional Alfv\'en waves is the Cowling resistivity, which describes the resistive dissipation of transversal currents to the ambient magnetic field. Since we are interested in the temporal evolution of waves, we assume that the perturbation in the magnetic field varies as $e^{ikz}$, where $k$ is the longitudinal wavenumber. As a result, Eq. (\ref{eq:3}) reduces to
\begin{equation}
b_{\varphi}=\frac{B_0}{ikv_A^2}\frac{d v_{\varphi}}{d t},
\label{eq:5}
\end{equation}
where $v_A=B_0/\sqrt{\mu_0\rho_0}$ is the Alfv\'en speed. By substituting the expression of $b_{\varphi}$ with the azimuthal component of the induction equation (Eq. \ref{eq:4}), we arrive at 
\begin{equation}
\frac{d^2 v_{\varphi}}{d t^2}+k^2\eta_C\frac{d v_{\varphi}}{d t}+k^2v_A^2v_{\varphi}=0.
\label{eq:6}
\end{equation}
The roots of the auxiliary equation of the above second-order differential equation are
\[
-\frac{k^2\eta_C}{2}\pm \frac{k}{2}\sqrt{k^2\eta_C^2-4v_A^2}.
\]
Since we aim to describe oscillatory motion, we need to impose the condition that $\eta_C<2v_A/k$, that is, the quantity under the square root is negative. This condition also means that waves oscillate if $k<2v_A/\eta_C$. The existence of  a cut-off wavenumber for propagating Alfv\'en waves was obtained earlier by Barc\'elo et al. (2010) and Zaqarashvili et al. (2011). Therefore, the solution of Eq. (\ref{eq:6}) can be written as
\begin{equation}
v_{\varphi}=Ae^{-k^2\eta_C t/2}\sin\left[kt\sqrt{v_A^2-\frac{k^2\eta_C^2}{4}}+\Phi\right],
\label{eq:7}
\end{equation}
where $\Phi$ is the phase of torsional Alfv\'en waves (undetermined from observations) and $A$ is the amplitude. The above relation describes an exponentially decaying sinusoidal signal, which is similar to the one observed and determined by Kohutova et al (2020). After comparing our solution with the form determined by observations, we arrive at the conclusion that the solution given by Eq. (\ref{eq:7}) describes the observed wave, provided that the amplitude of waves is 41 km s$^{-1}$ and that the damping time, $\tau$, and the period of waves, $P$, are given by
\[
\tau= \frac{2}{k^2\eta_C^2}=136 \;\;s, \quad P=\frac{2\pi}{k\sqrt{v_A^2-k^2\eta_C^2/4}}=89 \;\; s.
\]
Combining the above two relations results in a formula that allows us to determine the Cowling resistivity based on observable temporal parameters (period and damping time)
\begin{equation}
\eta_C=\frac{2v_A^2P^2\tau}{4\pi^2 \tau^2+P^2}.
\label{eq:8}
\end{equation}
The value of the Alfv\'en speed was determined using observations and employs the value of 140 km s$^{-1}$. Inserting all of the observed values into Eq. (\ref{eq:8}) we obtain that $\eta_C=5.73\times 10^{10}$ m$^2$ s$^{-1}$. This value is compared to the theoretical expression of $\eta_C$, which is given by Eq. (\ref{eq:2.1}), to determine the quantity $\xi_n$ that provides information about the ionisation state of the prominence plasma where torsional Alfv\'en waves were observed. After simple calculations, we affirm that the number density of neutrals at $T=10^4$ K is approximately $5.08\times 10^{16}$ m$^{-3}$, which is 4.23 times larger than the number density of neutrals predicted by the FAL-3 solar atmospheric model. Interestingly, this value is almost identical with the VAL-C model prediction (Vernazza et al. 1981).

Although the observations made by Kohutova et al. (2020) constitute the first observation of torsional Alfv\'en waves in the solar prominence plasma, the observed values for damping time, period, amplitude, and phase speed do not contain the error range that usually comes with observationally determined parameters. That is why we would like to investigate what effect a 10\% error range in physical parameters would have on the determined value of neutral number density.  After simple estimations, it turns out that the hypothetised change in measured wave parameters leads to a neutral number density of $(5.08\pm 0.05)\times 10^{16}$ m$^{-3}$  in the predicted number of neutrals, meaning that observational errors have a less important effect on the diagnostics of the ionisation degree of the plasma.

Finally, with the determined Cowling resistivity, we can define the wavelength range for which the present analysis is valid. In using the validity condition of oscillatory motion, it is easy to show that our analysis is valid provided the wavelength of waves is larger than 1.28 Mm or $k<4.9\times 10^{-6}$ m$^{-1}$ (wavenumbers and wavelengths that are currently observed in solar prominences, see, e.g. Arregui et al. 2012) . However, this wavelength is larger than the gravitational scale-height corresponding to $T=10^4$ K, meaning that for a more accurate description of torsional Alfv\'en waves observed by Kohutova et. al. (2020), one needs to take gravitational stratification into account. Furthermore, an in-depth study would benefit from a comprehensive analysis of dominant dissipative effects (other than only Cowling resistivity) or the role of phase mixing or resonant absorption, which could also be candidates for explaining the observed wave damping (similar to Ballai 2003 or Arregui and Ballester 2011).




%
%

\end{document}